\documentclass[12pt]{article}
\usepackage[cp866]{inputenc}
\usepackage{textcomp}
\usepackage{amsfonts,amssymb,amsmath,fullpage,pst-node,rotating,graphics,epsfig,subfigure,amsthm}
\usepackage{multirow}
\usepackage{authblk}

\newcommand{\ri}{{\rm i}\,}

\newcommand{\cri}{{\rm crit}}

\newcommand{\cM}{{\cal M}}
\newcommand{\rme}{{\rm e}}
\newcommand{\rr}{{\rm r}}

\newcommand{\bbf}{{\bf f}}

\newcommand{\bv}{{\bf v}}

\newcommand{\bom}{\mbox{\boldmath${\Omega}$}}

\newtheorem{theorem}{Theorem}
\newtheorem{corollary}{Corollary}
\newtheorem{remark}{Remark}
\def\al{ {\it et al}.}
\begin{document}

\title{Stability of convective rolls in a horizontal layer rotating about an
inclined axis}

\author{Olga Podvigina\\
Institute of Earthquake Prediction Theory\\
and Mathematical Geophysics\\
84/32 Profsoyuznaya St., 117997 Moscow, Russian Federation}

\maketitle

\begin{abstract}
We present three results on stability of rolls in Boussinesq convection
in a plane horizontal layer with rigid boundaries that is rotating about
an inclined axis with the angular velocity $\bom=(\Omega_1,\Omega_2,\Omega_3)$.

$i.$ We call the {\it full} problem the set of equations governing
the temporal behaviour of the flow and temperature for an arbitrary $\bom$,
and by the {\it reduced} problem the set of equations for the angular velocity
$(\Omega_1,0,\Omega_3)$. Here $x,y$ are horizontal Cartesian coordinates
in the layer and $z$ is the vertical one. We prove that a $y$-independent
solution to one of the two problems is also a solution to the second one.

$ii.$ We calculate the critical Rayleigh number for the monotonic
onset of convection. The instability mode in the form of rolls (a
flow independent of a horizontal direction) is assumed.
Let $\beta$ be the angle between the horizontal projection of $\bom$ and
the rolls axes. We show that $\beta=0$ for the least stable mode.
When the axis of rotation is horizontal, this is proven analytically, and
for $\Omega_3\ne0$, the result is obtained numerically. Taking $i$ into
account, we conclude that the critical Rayleigh number for the onset
of convection is independent of $\Omega_1$ and $\Omega_2$ and
the emerging flow are rolls with axis aligned with
the horizontal component of the rotation vector.

$iii.$ We study the behaviour of convective flows by integrating numerically
the three-dimensional equations of convection for $\bom=(0,\Omega_2,\Omega_3)$
and a range of the Rayleigh
numbers, other parameters of the problem being fixed. We assume square
horizontal periodicity cells, whose sides are equal to the period of the most
unstable mode. The computations indicate that, in general, in the nonlinear
regime convective rolls become more stable as $\Omega_2$ increases.
Namely, on increasing $\Omega_2$, the interval of the Rayleigh numbers
for which convective rolls are stable increases.
\end{abstract}
\maketitle


\section{Introduction}

Instability of thermal convection is a classical problem of the fluid dynamics
theory. Convective motion in a rotating layer is relevant for many geophysical
and astrophysical applications, such as the motion of the melt in the Earth's
outer core or of plasma in the convective layer of the Sun. Investigating
convective flows is also important for understanding the generation
of the magnetic field of stars and planets. For instance, according
to the present-day paradigm, the geomagnetic field is generated
by convective flows in the outer core (although this process is much more
complex involving many physical processes, such as phase transitions,
sedimentation, and chemical reactions). Since the pioneering work \cite{gla},
this hypothesis has been substantiated by many numerical studies
\cite{rob,chr}.

We consider the Boussinesq fluid in a plane horizontal layer heated
from below and rotating about an inclined axis.
Rigid horizontal boundaries held at constant temperatures are assumed.
The system in a dimensionless form is
characterized by the following parameters: the Rayleigh number, $R$, (measuring
the amplitude of thermal buoyancy forces), the Prandtl number, $P$, (the ratio
of kinematic viscosity to thermal diffusivity),
the square root of the Taylor number, $\tau=Ta^{1/2}$,
(proportional to the speed of rotation) and the unit vector along the direction
of the rotation axis, ${\bf e}_{\rr}$.

For small Rayleigh numbers, the fluid is not moving and the heat is transported
by thermal diffusion only. On increasing the Rayleigh number, a fluid motion
sets in. In the absence of rotation the instability is monotonic and the
emerging flow has the form of two-dimensional rolls \cite{chan}.
On increasing $R$ further, the rolls become unstable,
giving rise to more complex three-dimensional structures. The stability of
rolls, possible bifurcations and the developing secondary, tertiary and
subsequent flows were extensively studied in literature, see, e.g.,
\cite{bus,get} and references therein.

Convection in a rotating layer was studied in detail only for the vertical
rotation axis. The instability is oscillatory for small $P$ and sufficiently
large $\tau$. Clune and Knobloch \cite{kno1} computed the boundary
in the $(P,\tau)$ plane between the monotonic and oscillatory onset
of convection.
In the weakly nonlinear stability limit for $P\to\infty$, rolls are unstable
at the onset of convection in a layer with stress-free boundaries for $Ta>2285$
(the so-called K\"uppers-Lortz instability) \cite{kl69}.
The stability of rolls in a layer with rigid boundaries was examined for several
finite $P$ by Clever and Busse \cite{cb79}. They found that the domain
of stable rolls in the $(k,R)$ plane (here $k$ is the wave number of rolls)
significantly reduces as the rotation rate augments, and
showed that a subcritical finite-amplitude convection is possible for $P<1$.
For more recent results see \cite{bpa} and \cite{kba02}, and references therein.

Less attention has been given to the rotating convection with an inclined
rotation vector. The linear stability of the trivial steady state
was studied numerically by Hathaway and co-authors \cite{htg80}. They
found that  the critical Rayleigh number for the mode comprised of rolls whose axis is
oriented ``north-south'', which corresponds to the alignment with the
horizontal component of the rotation vector, is always smaller than for the
``east-west'' rolls. Similarly, in nonlinear numerical
simulations of Hathaway and Somerville \cite{hs83} it was also noted that the presence of the horizontal
component of the rotation vector ``elongates the convection cells in a
north-south direction''.
The limit of fast rotation (i.e., large Taylor number) was investigated by
Julien and Knobloch \cite{jk98}. Assuming that for $Ta\to\infty$ the critical
Rayleigh number and the critical wave number scale as $Ta^{2/3}$ and $Ta^{1/6}$, respectively,
 they found the critical values both for steady and oscillatory onset
of convection for the modes comprised of ``north-south'' and ``east-west'' rolls.
For no-slip horizontal boundaries the asymptotic regime is
reached at $Ta\approx10^{10}$. The most unstable rolls are oriented ``north-south'',
in agreement with earlier studies.
Similar preference to this orientation persists in the nonlinear regime.

For geophysical and astrophysical applications, the influence of other physical
processes
is also important. The combined effect of rotation and an imposed magnetic
field on the linear stability was investigated by Eltayeb \cite{el72,el75},
who studied analytically the asymptotics of the critical Rayleigh number
in the limit of large $Ta$ and $Q$ for the vertical or horizontal rotation axis
(here $Q$ is the Chandrasekhar number measuring the strength of the imposed
magnetic field). Zhang\al\ \cite{zhan3} computed the critical Rayleigh numbers
as a function of $\tau$ for several fixed values of $P$, $Q$ and the magnetic
Prandtl number $P_m$. The regions on the $(P,\tau)$ plane where rolls are stable
at the onset of rotating magnetoconvection were identified numerically
in \cite{op10} for several $P_m$ and $Q$. The combined effect of shear and
rotation about an inclined axis was studied in \cite{sow}.

Here, we study the stability of rolls in a layer with rigid horizontal boundaries
rotating about an inclined axis with the angular velocity $(\Omega_1,\Omega_2,\Omega_3)=\tau{\bf e}_{\rr}$.
We introduce a so-called reduced problem, where it is assumed $\Omega_2=0$.
We prove that a solution to the reduced problem that is independent of $y$
is also a solution to the full problem for an arbitrary $\Omega_2$ and the same
$\Omega_1$ and $\Omega_3$. (Here $x,y$ are horizontal Cartesian coordinates
in the layer and $z$ is the vertical one). In particular, convective rolls
parallel to the $y$ axis that solve the reduced problem remain a solution
to the full problem for any $\Omega_2$.

To calculate the critical Rayleigh number for the onset of convection,
we follow the approach that Pellew and Southwell \cite{ps40} employed
for calculating the critical values for a non-rotating layer; it was also
used in \cite{kno1,zhan3,op10,zhan1} for a layer rotating about a vertical
axis or/and with imposed magnetic field.
An instability mode of the form of two-dimensional convective rolls is
assumed to be a finite sum of products of trigonometric functions and
exponential functions with unknown coefficients. Substituting the sum into
the equations of convection and boundary conditions yields a nonlinear
system of equations. Solving them numerically, we determine the critical
Rayleigh number for a given $\bom=(\Omega_1,\Omega_2,\Omega_3)$.
(For monotonic instability the critical number is independent of $P$.)
The critical number also depends on the period of the rolls in the horizontal
direction characterized by the wave number $k$ and the angle $\beta$ between
the axis of rolls and the horizontal component of $\bom$; we express this
explicitly as $R_c(\bom,k,\beta)$. Minimizing over $\beta$
and $k$, we determine the critical Rayleigh number and critical $k$ for the
prescribed $\bom$.

Our results reveal that the minimum of $R_c(\bom,k,\beta)$ is achieved for $\beta=0$.
For a horizontal rotation axis ($\Omega_3=0$), we prove this analytically.
Hence, the critical Rayleigh number and the critical wave number are the same
as for the non-rotating layer. For $\Omega_3\ne0$, numerical results indicate
that the minimum over $\beta$ is also achieved for $\beta=0$, implying that
the critical Rayleigh number is the same as in the well-studied
problem of the onset of convection in the layer rotating about a vertical axis.

Hence, the convective rolls bifurcating in the monotonic
instability of the trivial solution (no motion) are the same for any values
of $\Omega_1$ and $\Omega_2$. Let us choose the coordinate system such that
$\Omega_1=0$. A natural question is how the stability of rolls
depends on $\Omega_2$ on increasing the Rayleigh number.
To answer it, we integrate numerically the
equations of convection with a horizontal square periodicity cell
for the wave number equal to that of the most unstable mode.
We compute bifurcation diagrams for a varying $R$, other parameters
of the problem being fixed. The computations show that the range of $R$
where the rolls are stable, in general, increases together with $\Omega_2$.

Hathaway and Somerville \cite{hs83} made a plausible statement based on previous studies
that ``a tilted rotation vector may give rise to a rich variety of
convective structures''. Our results indicate that near the onset of convection
this statement is wrong, because the change of rotation vector from
$(0,0,\Omega_3)$ to $(0,\Omega_2,\Omega_3)$ does not modify the convective
flow emerging at the onset of convection. Moreover, non-vanishing $\Omega_2$
makes the emerging rolls more stable, hence the bifurcations to more complex
patterns take place at larger values of $R$, compare to $\Omega_2=0$.

The paper is organized as follows. In section \ref{sec2}, we recall the basic
equations of convection and the boundary conditions.
In section \ref{sec0}, we prove that the $y$-independent solutions for
the full and reduced problem coincide. In section \ref{sec3}, we recall
the results of \cite{op10} for plane convective layer with a vertical
rotation axis. In section \ref{sec4}, we study the onset of convection
for a horizontal axis of rotation
and in section \ref{sec5} for an inclined one.
Nonlinear convection is considered in section \ref{sec6}.  In conclusion, we briefly
summarize results and indicate directions for further studies.

\section{Equations and parameters}\label{sec2}

The Boussinesq convection obeys the Navier--Stokes equation
\begin{equation}\label{nst}
{\partial{\bf v}\over\partial t}={\bf v}\times(\nabla\times{\bf v})
+P\tau{\bf v}\times{\bf e}_{\rr}
+P\Delta{\bf v}+PR\theta{\bf e}_z-\nabla p
\end{equation}
and the heat transfer equation
\begin{equation}\label{htr}
{\partial\theta\over\partial t}=-({\bf v}\cdot\nabla)\theta+v_z+\Delta\theta,
\end{equation}
where ${\bf e}_{\rr}$ is the unit vector parallel to the rotation axis.
The fluid is incompressible:
\begin{equation}\label{inc0}
\nabla\cdot{\bf v}=0.
\end{equation}
Here $\bf v$ denotes the flow velocity, $\theta$
the difference between the flow temperature and the linear temperature profile
and $p$ is the modified pressure.

The equations involve the following dimensionless parameters:
$$
P={\nu\over\kappa},\
R={\alpha_v g d^3\over\nu\kappa}\delta T,\
\tau={2\Omega d^2\over\nu}.
$$
Here $d$ is the width of the layer, $\kappa$ thermal diffusivity,
$\rho$ mass density, $\nu$ kinematic viscosity, $\alpha_v$ volumetric expansion
coefficient, $g$ gravity acceleration and $\delta T$ the temperature
difference between the lower and upper boundaries of the layer.
Equations (\ref{nst})--(\ref{inc0}) can be expressed as
\begin{equation}\label{neq}
{\partial{\bf w}\over\partial t}=L{\bf w}+N({\bf w},{\bf w}),
\end{equation}
where
\begin{equation}\label{4com}
{\bf w}=({\bf v},\theta)
\end{equation}
is a 4-component vector field and
\begin{equation}\label{oplr}
L{\bf w}=
\left(
\begin{array}{l}
P\Delta{\bf v}+PR\theta{\bf e}_z-\nabla p'+
P\tau{\bf v}\times{\bf e}_\rr\\
v_z+\Delta\theta
\end{array}
\right),
\end{equation}
where $\nabla p'$ is determined from the condition that the sum in the r.h.s.
of the upper equation is solenoidal.
The expression for the bilinear function $N(\cdot,\cdot)$ can be found in \cite{op10}.

The flow satisfies the no-slip condition on the horizontal boundaries,
which are kept at constant temperatures:
\begin{equation}\label{bou1}
v_x=v_y=v_z=0,\quad\theta=0\quad\hbox{ at }z=\pm{1\over 2}.
\end{equation}

The system (\ref{neq})--(\ref{bou1}) has a trivial steady state ${\bf w}=0$,
whose stability is determined by the dominant
eigenvalue of (\ref{oplr}). The steady state
becomes unstable when the eigenvalue crosses the imaginary axis.

\section{Persistence of a solution to the equations of convection
for a modified angular velocity}\label{sec0}

In the Navier--Stokes equation, the product
$\tau{\bf e}_{\rr}\equiv\bom=(\Omega_1,\Omega_2,\Omega_3)$
expresses the angular velocity of the convective layer. In this notation,
the equation (\ref{nst}) can be re-written as follows:
\begin{equation}\label{nst1}
{\partial{\bf v}\over\partial t}={\bf v}\times(\nabla\times{\bf v})
+P{\bf v}\times{\bom}
+P\Delta{\bf v}+PR\theta{\bf e}_z-\nabla p
\end{equation}
In this section we show that $y$-independent solutions to the equations
(\ref{nst1}),(\ref{htr}),(\ref{inc0}) for $\bom=(\Omega_1,0,\Omega_3)$
remain solutions to these equations for $\bom=(\Omega_1,\Omega_2,\Omega_3)$
and an arbitrary $\Omega_2$.

\begin{theorem}\label{th0}
Suppose ${\bf w}(t)=({\bf v}(t),\theta(t))$ is a solution to
(\ref{nst1}),(\ref{htr}),(\ref{inc0}) for
\begin{equation}\label{om0}
\bom=(\Omega_1,0,\Omega_3),
\end{equation}
such that both ${\bf v}(t)$ and $\theta(t)$ depend on $x$ and $z$ only.
Then ${\bf w}(t)$ is a solution to (\ref{nst1}),(\ref{htr}),(\ref{inc0}) (upon
a suitable modification of the pressure $p$) for
\begin{equation}\label{om1}
\bom=(\Omega_1,\Omega_2,\Omega_3)
\end{equation}
and any $\Omega_2$.
\end{theorem}

\proof
Denote $\bom_{2}=(0,\Omega_2,0)$ and $\bom_{13}=(\Omega_1,0,\Omega_3)$ .
According to the statement of the theorem
the fields $({\bf v}(t),\theta(t))$ satisfy the equations
(\ref{nst1}),(\ref{htr}) with $\bom=\bom_{13}$. Let $\tilde p(x,z,t)$
be a solution to the equations
$$\partial\tilde p(x,z,t)/\partial x=-v_z,\
\partial\tilde p(x,z,t)/\partial z=v_x.$$
The incompressibility of $\bf v$ and its independence of $y$ imply
that such $\tilde p(x,z,t)$ exists. Hence,
$${\bf v}\times{\bom}_2=
\left(
\begin{array}{l}
-\Omega_2v_z\\
0\\
\Omega_2v_x
\end{array}
\right)
=\Omega_2\nabla\tilde p.
$$
Substitution of $\bom=\bom_2+\bom_{13}$ into (\ref{nst1}) implies that the r.h.s.
of the equation takes the form
$$
\renewcommand{\arraystretch}{1.8}
\begin{array}{l}
{\bf v}\times(\nabla\times{\bf v})
+P{\bf v}\times(\bom_2+\bom_{13})
+P\Delta{\bf v}+PR\theta{\bf e}_z-\nabla p=\\
{\bf v}\times(\nabla\times{\bf v})
+P{\bf v}\times\bom_{13}+P{\bf v}\times\bom_2
+P\Delta{\bf v}+PR\theta{\bf e}_z-\nabla p=\\
{\bf v}\times(\nabla\times{\bf v})
+P{\bf v}\times\bom_{13}
+P\Delta{\bf v}+PR\theta{\bf e}_z-\nabla p+P\Omega_2\nabla\tilde p.
\end{array}
$$
Consequently,
the modification of $\bom$ changes the pressure only. The theorem is proven.
\qed

\begin{remark}\label{rm0}
Note that the theorem is true for any boundary conditions on $\bf v$ and $\theta$,
not only for (\ref{bou1}) employed in this paper.
\end{remark}

\section{Linear stability of the trivial steady state, a vertical
rotation axis}\label{sec3}

For the sake of completeness of the presentation, here we consider the onset
of convection in a horizontal layer rotating about a vertical axis.
Namely, the results of subsection III.A of \cite{op10} are formulated as
a theorem providing the equations for determining the critical
Rayleigh number for the monotonic instability.

As usual in the study of the onset of convection, we assume that the
critical mode, which is a solution to the equation
\begin{equation}\label{eigrr}
L{\bf w}=0,
\end{equation}
depends on two variables, $x$ and $z$.
The mode (\ref{4com}) takes the form
\begin{equation}\label{seriw}
{\bf w}=\sum_{j=1}^3\sum_{l=1}^8A_{jl}{\bf f}_j(k,\gamma_l),
\end{equation}
where the 4-component vector fields ${\bf f}_j$ are as follows:
\begin{equation}\label{bas1}
{\bf f}_1(k,\gamma)=\left(
\begin{array}{c}
-\displaystyle\frac{\gamma}{k}{\rm e}^{\gamma z}\sin kx\\
0\\
{\rm e}^{\gamma z}\cos kx\\
0
\end{array}
\right),\
{\bf f}_2(k,\gamma)=\left(
\begin{array}{c}
0\\
{\rm e}^{\gamma z}\sin kx\\
0\\
0
\end{array}
\right),\
{\bf f}_3(k,\gamma)=\left(
\begin{array}{c}
0\\
0\\
0\\
{\rm e}^{\gamma z}\cos kx
\end{array}
\right).
\end{equation}

As noted by Chandrasekhar \cite{chan}, an instability mode of the trivial
steady state can be {\it even} or {\it odd}. Even modes satisfy the relations
\begin{equation}\label{even}
\bv(-x,-y,-z)=\bv(x,y,z),\ \theta(-x,-y,-z)=\theta(x,y,z),
\end{equation}
and odd modes have the opposite parity. For convection with rotation
about the vertical axis, the critical Rayleigh number
for the odd mode is substantially larger than for the even mode.
We have checked that this remains true for the inclined
rotation axis. Thus, only even modes are important near
the onset of convection and only even modes are discussed in this paper.

The main result concerning the monotonic onset of convection can be
summarized as the following theorem:

\begin{theorem}\label{th1}
Suppose a vector field (\ref{seriw})--(\ref{even})
is a solution to equations (\ref{eigrr}),(\ref{oplr}),(\ref{inc0}),(\ref{bou1})
for $\rme_{\rr}=(0,0,1)$. Then (after an appropriate permutation)
the eight exponents $\gamma_s$ in (\ref{seriw}) satisfy the relations
\begin{equation}\label{gamar}
\gamma_1=k,\ \gamma_2=-k,\ \gamma_{2s+1}=-\gamma_{2s+2},\ s=1,2,3.
\end{equation}
For $s=3,...,8$ the exponents are roots of the equation
\begin{equation}\label{detr00}
(\gamma^2-k^2)^3+\tau^2\gamma^2+Rk^2=0
\end{equation}
and
\begin{equation}\label{matbr0}
\det\left(
\renewcommand{\arraystretch}{2.3}
\begin{array}{cccc}
0&1&1&1\\
0&\gamma_3\tanh\gamma_3/2&\gamma_5\tanh\gamma_5/2&\gamma_7\tanh\gamma_7/2\\
\tau&\displaystyle\frac{1}{\gamma_3^2-k^2}&
\displaystyle\frac{1}{\gamma_5^2-k^2}&
\displaystyle\frac{1}{\gamma_7^2-k^2}\\
-Rk\tanh k/2&\displaystyle\frac{\tau\gamma_3\tanh\gamma_3/2}{\gamma_3^2-k^2}&
\displaystyle\frac{\tau\gamma_5\tanh\gamma_5/2}{\gamma_5^2-k^2}&
\displaystyle\frac{\tau\gamma_7\tanh\gamma_7/2}{\gamma_7^2-k^2}
\end{array}
\right)=0.
\end{equation}
\end{theorem}

\proof
Denote the three first components of $\bbf_j$ by $\bv_j$ and the fourth
component by $\theta_j$. For $\gamma=\pm k$,
$$
R(P\Delta{\bf v}_2+P\tau{\bf v}_2\times{\bf e}_\rr)\mp\tau PR\theta_3{\bf e}_z=
PR\tau
\left(
\begin{array}{c}
{\rm e}^{\gamma z}\sin kx\\
0\\
0
\end{array}
\right)\mp
PR\tau
\left(
\begin{array}{c}
0\\
0\\
{\rm e}^{\gamma z}\cos kx
\end{array}
\right)=\nabla p',
$$
where $p'=-PR\tau{\rm e}^{\gamma z}\cos kx/k$.
Since $\theta_2=0$, $\bv_3=0$ and $\Delta\theta_3=0$, we have that
\begin{equation}\label{detr10}
L[R{\bf f}_2(k,\gamma)\mp\tau{\bf f}_3(k,\gamma)]=0.
\end{equation}

For $\gamma\ne\pm k$,
$$
P\Delta{\bf v}_1+P\tau{\bf v}_1\times{\bf e}_\rr=
P(\gamma^2-k^2)\bv_1+
P\tau
\left(
\begin{array}{c}
0\\
\displaystyle\frac{\gamma}{k}{\rm e}^{\gamma z}\sin kx\\
0
\end{array}
\right)=P(\gamma^2-k^2)\bv_1+{P\tau\gamma\over k}\bv_2,
$$
$$
\renewcommand{\arraystretch}{1.0}
\begin{array}{l}
P\Delta{\bf v}_2+P\tau{\bf v}_2\times{\bf e}_\rr=P(\gamma^2-k^2)\bv_2+
P\tau
\left(
\begin{array}{c}
{\rm e}^{\gamma z}\sin kx\\
0\\
0
\end{array}
\right)=\\
P(\gamma^2-k^2)\bv_2
-\displaystyle\frac{P\tau\gamma k}{\gamma^2-k^2}
\left(
\begin{array}{c}
-\displaystyle\frac{\gamma}{k}{\rm e}^{\gamma z}\sin kx\\
0\\
{\rm e}^{\gamma z}\cos kx
\end{array}
\right)+
\displaystyle\frac{P\tau\gamma k}{\gamma^2-k^2}
\left(
\begin{array}{c}
-\displaystyle\frac{k}{\gamma}{k}{\rm e}^{\gamma z}\sin kx\\
0\\
{\rm e}^{\gamma z}\cos kx
\end{array}
\right)=\\
P(\gamma^2-k^2)\bv_2
-\displaystyle\frac{P\tau\gamma k}{\gamma^2-k^2}\bv_1+
\displaystyle\frac{P\tau k}{\gamma^2-k^2}\nabla{\rm e}^{\gamma z}\cos kx
\end{array}
$$
and
$$
\begin{array}{l}
PR\theta_3{\bf e}_z=
PR\left(
\begin{array}{c}
0\\
0\\
{\rm e}^{\gamma z}\cos kx
\end{array}
\right)=\\
-\displaystyle\frac{PRk^2}{\gamma^2-k^2}
\left(
\begin{array}{c}
-\displaystyle\frac{\gamma}{k}{\rm e}^{\gamma z}\sin kx\\
0\\
{\rm e}^{\gamma z}\cos kx
\end{array}
\right)+
\displaystyle\frac{PR\gamma^2}{\gamma^2-k^2}
\left(
\begin{array}{c}
-\displaystyle\frac{k}{\gamma}{k}{\rm e}^{\gamma z}\sin kx\\
0\\
{\rm e}^{\gamma z}\cos kx
\end{array}
\right)=
-\displaystyle\frac{PRk^2}{\gamma^2-k^2}\bv_1+
\displaystyle\frac{PR\gamma}{\gamma^2-k^2}\nabla{\rm e}^{\gamma z}\cos kx
.
\end{array}
$$
Hence,
\begin{equation}\label{actL}
\renewcommand{\arraystretch}{1.5}
\begin{array}{l}
L\bbf_1=P(\gamma^2-k^2)\bbf_1+
\displaystyle\frac{\tau\gamma P}{k}\bbf_2+\bbf_3,\\
L\bbf_2=
-\displaystyle\frac{\tau\gamma kP}{\gamma^2-k^2}\bbf_1+P(\gamma^2-k^2)\bbf_2,\\
L\bbf_3=-\displaystyle\frac{PRk^2}{\gamma^2-k^2}\bbf_1+(\gamma^2-k^2)\bbf_3.
\end{array}
\end{equation}

Therefore, the matrix of $L$ acting on $F(k,\gamma)$ is
\begin{equation}\label{matL0}
\left(
\renewcommand{\arraystretch}{1.5}
\begin{array}{ccc}
P(\gamma^2-k^2)&-\displaystyle\frac{\tau\gamma kP}{\gamma^2-k^2}&
-\displaystyle\frac{PRk^2}{\gamma^2-k^2}\\
\displaystyle\frac{\tau\gamma P}{k}&P(\gamma^2-k^2)&0\\[1ex]
1&0&(\gamma^2-k^2)
\end{array}
\right).
\end{equation}
The operator $L$ possesses a non-trivial kernel in $F(k,\gamma)$
when the determinant of (\ref{matL0}), which is the product of
the l.h.s. of (\ref{detr00}) and  $P^2$,
vanishes. Hence, (\ref{eigrr}) has a non-trivial solution whenever
(\ref{detr00}) holds. Therefore, a solution to (\ref{eigrr}) takes the form
(\ref{seriw}), where the exponents $\gamma$ satisfy (\ref{detr00}).

By virtue of (\ref{detr10}) and (\ref{actL}), if (\ref{seriw})
is a solution to (\ref{eigrr}), the coefficients $A_{jl}$ satisfy the relations
\begin{equation}\label{coeA}
\renewcommand{\arraystretch}{2.3}
\begin{array}{l}
A_{11}=A_{12}=0,\ A_{21}=-\displaystyle{R\over\tau}A_{31},\ A_{22}=\displaystyle{R\over\tau}A_{32},\\
A_{2l}=-\displaystyle{\tau \gamma_l\over k(\gamma_l^2-k^2)}A_{1l},\
A_{3l}=-\displaystyle{1\over(\gamma_l^2-k^2)}A_{1l},\ l=3,\ldots,8.
\end{array}
\end{equation}

We seek an eigenmode (\ref{seriw}) subject to the boundary conditions
(\ref{bou1}). Substituting (\ref{seriw}) into (\ref{bou1})
yields a system of linear homogeneous equations in $A_{jl}$, $l=1,\ldots,8$:
\begin{equation}\label{sysA}
\sum_{l=1}^8{\rm e}^{\pm\gamma_l/2}A_{jl}=0,\ j=1,2,3,\quad
\sum_{l=1}^8{\gamma_l\over k}{\rm e}^{\pm\gamma_l/2}A_{1l}=0.
\end{equation}
For an even eigenmode (\ref{seriw}),(\ref{bas1}) if $\gamma_l$ satisfies (\ref{gamar}),
then $A_{2,2s+1}=-A_{2,2s+2}$ and $A_{j,2s+1}=A_{j,2s+2}$, $s=1,2,3$, $j=1,3$.
Therefore, (\ref{sysA}) is equivalent to the system of four equations:
\begin{equation}\label{sysAm}
\renewcommand{\arraystretch}{2.3}
\begin{array}{l}
A_{11}({\rm e}^{\gamma_1/2}+{\rm e}^{-\gamma_1/2})+
A_{13}({\rm e}^{\gamma_3/2}+{\rm e}^{-\gamma_3/2})+
A_{15}({\rm e}^{\gamma_5/2}+{\rm e}^{-\gamma_5/2})+
A_{17}({\rm e}^{\gamma_7/2}+{\rm e}^{-\gamma_7/2})=0\\
A_{21}({\rm e}^{\gamma_1/2}-{\rm e}^{-\gamma_1/2})+
A_{23}({\rm e}^{\gamma_3/2}-{\rm e}^{-\gamma_3/2})+
A_{25}({\rm e}^{\gamma_5/2}-{\rm e}^{-\gamma_5/2})+
A_{27}({\rm e}^{\gamma_7/2}-{\rm e}^{-\gamma_7/2})=0\\
A_{31}({\rm e}^{\gamma_1/2}+{\rm e}^{-\gamma_1/2})+
A_{33}({\rm e}^{\gamma_3/2}+{\rm e}^{-\gamma_3/2})+
A_{35}({\rm e}^{\gamma_5/2}+{\rm e}^{-\gamma_5/2})+
A_{37}({\rm e}^{\gamma_7/2}+{\rm e}^{-\gamma_7/2})=0\\
A_{11}\gamma_1({\rm e}^{\gamma_1/2}-{\rm e}^{-\gamma_1/2})+
A_{13}\gamma_3({\rm e}^{\gamma_3/2}-{\rm e}^{-\gamma_3/2})+
A_{15}\gamma_5({\rm e}^{\gamma_5/2}-{\rm e}^{-\gamma_5/2})+
A_{17}\gamma_7({\rm e}^{\gamma_7/2}-{\rm e}^{-\gamma_7/2})=0
\end{array}
\end{equation}
Substituting (\ref{coeA}) into (\ref{sysAm}), we obtain equations involving
4 unknown quantities, $A_{21}$, $A_{13}$, $A_{15}$ and $A_{17}$:
\begin{equation}\label{sysAmm}
\renewcommand{\arraystretch}{2.5}
\begin{array}{l}
A_{13}\cosh\gamma_3/2+A_{15}\cosh\gamma_5/2+A_{17}\cosh\gamma_7/2=0\\
A_{21}\sinh\gamma_1
-A_{13}\displaystyle{\tau \gamma_3\sinh\gamma_3/2\over k(\gamma_3^2-k^2)}
-A_{15}\displaystyle{\tau \gamma_5\sinh\gamma_5/2\over k(\gamma_5^2-k^2)}
-A_{17}\displaystyle{\tau \gamma_7\sinh\gamma_7/2\over k(\gamma_7^2-k^2)}=0\\
-A_{21}\displaystyle{\tau\over R}\cosh\gamma_1
-A_{13}\displaystyle{\cosh\gamma_3/2\over (\gamma_3^2-k^2)}
-A_{15}\displaystyle{\cosh\gamma_5/2\over (\gamma_5^2-k^2)}
-A_{17}\displaystyle{\cosh\gamma_7/2\over (\gamma_7^2-k^2)}=0\\
A_{13}\gamma_3\sinh\gamma_3/2+A_{15}\gamma_5\sinh\gamma_5/2+
A_{17}\gamma_7\sinh\gamma_7/2=0.
\end{array}
\end{equation}
It has a non-zero solution, if the determinant of the respective $4\times4$
matrix vanishes. Multiplying the second row of the matrix of the linear
system (\ref{sysAmm}) by $-k$, the third one by $-1$,
the first colomn by $R$, permuting the second and the last rows, and
dividing the $j$-th colomn by $\cosh\gamma_{2j-1}/2$
we obtain the matrix in the l.h.s. of (\ref{matbr0}).
\qed

For a given $\tau$ and $k$, the critical value $R_c(\tau,k)$ is a solution
to (\ref{matbr0}), where $\gamma_l$, $l=3,5,7$, are defined by (\ref{detr00}).
Minimizing $R_c(\tau,k)$ over $k$, we determine the critical Rayleigh number $R^{\cri}(\tau)$
and the critical wave number $k^{\cri}(\tau)$ for the monotonic instability of the trivial
steady state.

\section{Linear stability of the trivial steady state, a horizontal
rotation axis}\label{sec4}

In a layer rotating about the vertical axis, the critical Rayleigh number
for an instability mode in the form of rolls (a two-component flow)
is independent of the direction of the rolls axes.
This is not the case if the rotation axis is horizontal or inclined.
As in section \ref{sec3}, we assume that a critical mode depends on $x$ and $z$.
We choose a coordinate system such that the $y$ axis is parallel to the rolls
axes. A unit vector in the direction of the rotation axis is
\begin{equation}\label{vech}
{\bf e}_\rr=\left(
\begin{array}{c}
\sin\beta\\
\cos\beta\\
0
\end{array}
\right),
\end{equation}
where $\beta$ is the angle between the rotation axis and rolls axes.
We call the instability mode {\it parallel} when $\beta=0$, and
{\it oblique} otherwise. The critical Rayleigh number depends on the
speed of rotation, on the wave number of rolls and on the angle $\beta$.

The mode is sought in the form of the ansatz (\ref{seriw}), where
\begin{equation}\label{bas2}
{\bf f}_1(k,\gamma)=\left(
\begin{array}{c}
-\displaystyle\frac{\gamma}{k}{\rm e}^{\gamma z}\sin kx\\
0\\
{\rm e}^{\gamma z}\cos kx\\
0
\end{array}
\right),\
{\bf f}_2(k,\gamma)=\left(
\begin{array}{c}
0\\
{\rm e}^{\gamma z}\cos kx\\
0\\
0
\end{array}
\right),\
{\bf f}_3(k,\gamma)=\left(
\begin{array}{c}
0\\
0\\
0\\
{\rm e}^{\gamma z}\cos kx
\end{array}
\right),
\end{equation}
i.e., while ${\bf f}_1(k,\gamma)$ and ${\bf f}_3(k,\gamma)$ are the same as in
(\ref{bas1}), in ${\bf f}_2(k,\gamma)$ we have replaced $\sin kx$ by
$\cos kx$. The equations determining the stability of the trivial state
with respect to convective rolls are derived in the following theorem.

\begin{theorem}\label{th2}
Suppose the vector field (\ref{seriw}),(\ref{bas2}),(\ref{even}) is
a solution to (\ref{eigrr}),(\ref{oplr}),(\ref{inc0}),(\ref{bou1}),
the direction of the rotation axis satisfying (\ref{vech}). Then (after an
appropriate permutation) the eight exponents $\gamma_s$ in (\ref{seriw})
satisfy
\begin{equation}\label{gamar2}
\gamma_1=k,\ \gamma_2=-k,\ \gamma_{2s+1}=-\gamma_{2s+2},\ s=1,2,3.
\end{equation}
For $s=3,...,8$ the exponents are roots of the equations
\begin{equation}\label{detr22}
(\gamma^2-k^2)^3+(R-\tau^2\sin^2\beta)k^2=0
\end{equation}
and
\begin{equation}\label{matbr2}
\det\left(
\renewcommand{\arraystretch}{2.2}
\begin{array}{ccc}
1&1&1\\
\gamma_3\tanh\gamma_3/2&\gamma_5\tanh\gamma_5/2&\gamma_7\tanh\gamma_7/2\\
\displaystyle\frac{1}{\gamma_3^2-k^2}&
\displaystyle\frac{1}{\gamma_5^2-k^2}&
\displaystyle\frac{1}{\gamma_7^2-k^2}
\end{array}
\right)=0.
\end{equation}
\end{theorem}

\proof
As in the proof of theorem \ref{th1}, the three first components of $\bbf_j$
are denoted by $\bv_j$ and the fourth one by $\theta_j$. For $\gamma=\pm k$,
$$
R(P\Delta{\bf v}_2+P\tau{\bf v}_2\times{\bf e}_\rr)
-\tau\sin\beta PR\theta_3{\bf e}_z=
PR\tau\sin\beta
\left(
\begin{array}{c}
0\\
0\\
{\rm e}^{\gamma z}\cos kx
\end{array}
\right)-
PR\tau\sin\beta
\left(
\begin{array}{c}
0\\
0\\
{\rm e}^{\gamma z}\cos kx
\end{array}
\right)=0.
$$
Therefore,
\begin{equation}\label{detr10h}
L[R{\bf f}_2(k,\gamma)-\tau\sin\beta{\bf f}_3(k,\gamma)]=0.
\end{equation}

For $\gamma\ne\pm k$,
$$
\begin{array}{l}
P\Delta{\bf v}_1+P\tau{\bf v}_1\times{\bf e}_\rr=
P(\gamma^2-k^2)\bv_1+
P\tau
\left(
\begin{array}{c}
-{\rm e}^{\gamma z}\cos kx\cos\beta\\
{\rm e}^{\gamma z}\cos kx\sin\beta\\
-\displaystyle\frac{\gamma}{k}{\rm e}^{\gamma z}\sin kx\cos\beta
\end{array}
\right)=\\
P(\gamma^2-k^2)\bv_1+P\tau\sin\beta\bv_2-
\displaystyle\frac{P\tau\cos\beta}{k}\nabla{\rm e}^{\gamma z}\sin kx
\end{array}
$$
$$
\renewcommand{\arraystretch}{1.0}
\begin{array}{l}
P\Delta{\bf v}_2+P\tau{\bf v}_2\times{\bf e}_\rr=P(\gamma^2-k^2)\bv_2+
P\tau
\left(
\begin{array}{c}
0\\
0\\
-{\rm e}^{\gamma z}\cos kx\sin\beta
\end{array}
\right)=\\
P(\gamma^2-k^2)\bv_2+
\displaystyle\frac{P\tau\sin\beta k^2}{\gamma^2-k^2}
\left(
\begin{array}{c}
-\displaystyle\frac{\gamma}{k}{\rm e}^{\gamma z}\sin kx\\
0\\
{\rm e}^{\gamma z}\cos kx
\end{array}
\right)-
\displaystyle\frac{P\tau\sin\beta\gamma^2}{\gamma^2-k^2}
\left(
\begin{array}{c}
-\displaystyle\frac{k}{\gamma}{k}{\rm e}^{\gamma z}\sin kx\\
0\\
{\rm e}^{\gamma z}\cos kx
\end{array}
\right)=\\
P(\gamma^2-k^2)\bv_2+
\displaystyle\frac{P\tau\sin\beta k^2}{\gamma^2-k^2}\bv_1-
\displaystyle\frac{P\tau\sin\beta\gamma}{\gamma^2-k^2}
\nabla{\rm e}^{\gamma z}\cos kx
\end{array}
$$
and
$$
\begin{array}{l}
PR\theta_3{\bf e}_z=
PR\left(
\begin{array}{c}
0\\
0\\
{\rm e}^{\gamma z}\cos kx
\end{array}
\right)=
-\displaystyle\frac{PRk^2}{\gamma^2-k^2}
\left(
\begin{array}{c}
-\displaystyle\frac{\gamma}{k}{\rm e}^{\gamma z}\sin kx\\
0\\
{\rm e}^{\gamma z}\cos kx
\end{array}
\right)+
\displaystyle\frac{PR\gamma^2}{\gamma^2-k^2}
\left(
\begin{array}{c}
-\displaystyle\frac{k}{\gamma}{k}{\rm e}^{\gamma z}\sin kx\\
0\\
{\rm e}^{\gamma z}\cos kx
\end{array}
\right)=\\
-\displaystyle\frac{PRk^2}{\gamma^2-k^2}\bv_1+
\displaystyle\frac{PR\gamma}{\gamma^2-k^2}\nabla{\rm e}^{\gamma z}\cos kx
.
\end{array}
$$
Hence,
\begin{equation}\label{actLh}
\renewcommand{\arraystretch}{1.8}
\begin{array}{l}
L\bbf_1=P(\gamma^2-k^2)\bbf_1+
P\tau\sin\beta\bbf_2+\bbf_3,\\
L\bbf_2=
\displaystyle\frac{P\tau\sin\beta k^2}{\gamma^2-k^2}\bbf_1+P(\gamma^2-k^2)\bbf_2,\\
L\bbf_3=-\displaystyle\frac{PRk^2}{\gamma^2-k^2}\bbf_1+(\gamma^2-k^2)\bbf_3.
\end{array}
\end{equation}

Therefore, for $\gamma\ne\pm k$, the matrix of $L$ acting on $F(k,\gamma)$ is
\begin{equation}\label{matL}
\left(
\renewcommand{\arraystretch}{1.5}
\begin{array}{ccc}
P(\gamma^2-k^2)&\displaystyle\frac{\tau\sin\beta k^2P}{\gamma^2-k^2}&
-\displaystyle\frac{PRk^2}{\gamma^2-k^2}\\
\tau\sin\beta P&P(\gamma^2-k^2)&0\\
1&0&(\gamma^2-k^2)
\end{array}
\right).
\end{equation}
The operator $L$ possesses a non-trivial kernel whenever the determinant
of (\ref{matL}) vanishes, i.e., when equality (\ref{detr22}) holds.

Due to (\ref{detr10h}) and (\ref{actLh}), (\ref{seriw}) is a solution
to (\ref{eigrr}) whenever the coefficients $A_{jl}$ satisfy
\begin{equation}\label{coeA1}
\renewcommand{\arraystretch}{2.3}
\begin{array}{l}
A_{11}=A_{12}=0,\ A_{31}=-\displaystyle{\tau\sin\beta\over R}A_{21},\
A_{32}=-\displaystyle{\tau\sin\beta\over R}A_{22},\\
A_{2l}=-\displaystyle{\tau\sin\beta\over(\gamma_l^2-k^2)}A_{1l},\
A_{3l}=-\displaystyle{1\over(\gamma_l^2-k^2)}A_{1l},\ l=3,\ldots,8.
\end{array}
\end{equation}

For an eigenmode (\ref{seriw}) subject to the boundary conditions
(\ref{bou1}) system of equations (\ref{sysA}) holds true.
By (\ref{coeA1}), system  (\ref{sysA}) can be regarded
as a linear system of four equations in $A_{21}$ and $A_{1l}$, $l=3,5,7$.
For an even mode the matrix of the system is
\begin{equation}\label{sysAmmh}
\renewcommand{\arraystretch}{2.5}
\left(
\begin{array}{cccc}
0 &\cosh\gamma_3/2 & \cosh\gamma_5/2 &\cosh\gamma_7/2\\
1 & -\displaystyle{\tau\sin\beta\cosh\gamma_3/2\over k(\gamma_3^2-k^2)}&
-\displaystyle{\tau\sin\beta \cosh\gamma_5/2\over k(\gamma_5^2-k^2)}&
-\displaystyle{\tau\sin\beta\cosh\gamma_7/2\over k(\gamma_7^2-k^2)}\\
\displaystyle{\tau\sin\beta\over R}&
-\displaystyle{\cosh\gamma_3/2\over (\gamma_3^2-k^2)}
&-\displaystyle{\cosh\gamma_5/2\over (\gamma_5^2-k^2)}&
-\displaystyle{\cosh\gamma_7/2\over (\gamma_7^2-k^2)}\\
0&\gamma_3\sinh\gamma_3/2&\gamma_5\sinh\gamma_5/2&\gamma_7\sinh\gamma_7/2
\end{array}
\right).
\end{equation}
The system has a non-trivial solution whenever the determinant of the matrix
vanishes. After dividing the columns 2, 3 and 4 by
$\cosh\gamma_3/2$, $\cosh\gamma_5/2$ and $\cosh\gamma_7/2$, respectively,
multiplying the first column and second and third rows by -1,
and permuting cyclically the last three rows,
matrix  (\ref{sysAmmh}) takes the form
\begin{equation}\label{matbr21}
\left(
\renewcommand{\arraystretch}{2.2}
\begin{array}{cccc}
0&1&1&1\\
0&\gamma_3\tanh\gamma_3/2&\gamma_5\tanh\gamma_5/2&\gamma_7\tanh\gamma_7/2\\
1&\displaystyle\frac{\tau\sin\beta}{\gamma_3^2-k^2}&
\displaystyle\frac{\tau\sin\beta}{\gamma_5^2-k^2}&
\displaystyle\frac{\tau\sin\beta}{\gamma_7^2-k^2}\\
\displaystyle{\tau\sin\beta\over R}&\displaystyle\frac{1}{\gamma_3^2-k^2}&
\displaystyle\frac{1}{\gamma_5^2-k^2}&
\displaystyle\frac{1}{\gamma_7^2-k^2}
\end{array}
\right).
\end{equation}
Multiplying the first column by $R\tau\sin\beta$, dividing the third row by
$\tau\sin\beta$ and subtracting the last row from the third one
we obtain the transformed matrix
\begin{equation}
\left(
\renewcommand{\arraystretch}{2.2}
\begin{array}{cccc}
0&1&1&1\\
0&\gamma_3\tanh\gamma_3/2&\gamma_5\tanh\gamma_5/2&\gamma_7\tanh\gamma_7/2\\
R-\tau^2\sin^2\beta&0&0&0\\
\tau^2\sin^2\beta&-\displaystyle\frac{1}{\gamma_3^2-k^2}&
-\displaystyle\frac{1}{\gamma_5^2-k^2}&
-\displaystyle\frac{1}{\gamma_7^2-k^2}
\end{array}
\right),
\end{equation}
whose determinant coincides with the l.h.s. of (\ref{matbr2})
up to the factor $R-\tau^2\sin^2\beta$.
\qed

\begin{corollary}
Denote by $R_c(\tau,k,\beta)$ the critical Rayleigh number for the onset
of convection in a layer rotating with the speed $\tau$ about
a horizontal axis for the mode of the form of rolls of wave number $k$, whose
axes are inclined by angle $\beta$ relative to the rotation axis. The theorem
implies that the critical Rayleigh number for such rolls satisfies the relation
$R_c(\tau,k,\beta)=R_c(0,k,0)+\tau^2\sin^2\beta$. Therefore,
the critical Rayleigh number for the parallel modes is smaller
than for the oblique ones. The former critical Rayleigh number, $R_c(0,k,0)$,
is known from the studies of the non-rotating convection.
\end{corollary}

\section{Linear stability of the trivial steady state, an inclined
rotation axis}\label{sec5}

Following the approach employed in sections \ref{sec3} and \ref{sec4}, here
we derive equations determining the critical Rayleigh number for the onset
of convection in a layer rotating about an inclined axis. An instability mode
is assumed to have the structure of two-dimensional rolls. We choose
the coordinate system where the $y$ axis is parallel to the rolls axes.
Let $\alpha$ be the angle between the rotation axis and the $z$ axis and
$\beta$ the angle between the rotation axis and the rolls axes. The unit
vector in the direction of the rotation axis is thus
\begin{equation}\label{inc}
{\bf e}_r=\left(
\begin{array}{c}
\sin\beta\\
\cos\beta\sin\alpha\\
\cos\alpha
\end{array}
\right).
\end{equation}
We again call {\it parallel} the instability mode for $\beta=0$.
For a fixed $\tau$ and $\alpha$, the critical Rayleigh
number depends on the wave number of rolls and on the angle $\beta$.

As in sections \ref{sec3} and \ref{sec4}, we introduce an $L$-invariant
subspace $F(k,\gamma)$. The subspace is spanned by the three vectors
(\ref{bas1}) and their images under the shift by $\pi/2$ in the $x$ direction.
This six-dimensional real subspace is regarded as a three-dimensional complex
one with the basis composed of 4-component vector fields
\begin{equation}\label{bas3}
\begin{array}{l}
{\bf f}_1(k,\gamma)=\left(
\begin{array}{c}
-\displaystyle\frac{\gamma}{k}{\rm e}^{\gamma z}\sin kx
-\ri\displaystyle\frac{\gamma}{k}{\rm e}^{\gamma z}\cos kx\\
0\\
{\rm e}^{\gamma z}\cos kx-\ri{\rm e}^{\gamma z}\sin kx\\
0
\end{array}
\right),\
{\bf f}_2(k,\gamma)=\left(
\begin{array}{c}
0\\
{\rm e}^{\gamma z}\sin kx+\ri{\rm e}^{\gamma z}\cos kx\\
0\\
0
\end{array}
\right),\\
{\bf f}_3(k,\gamma)=\left(
\begin{array}{c}
0\\
0\\
0\\
{\rm e}^{\gamma z}\cos kx-\ri{\rm e}^{\gamma z}\sin kx
\end{array}
\right).
\end{array}
\end{equation}
The instability mode is sought in the form (\ref{seriw}).
We derive the equation determining the stability of convective rolls
in the following theorem.

\begin{theorem}\label{th3}
Suppose a vector field (\ref{seriw}),(\ref{bas3}),(\ref{even})
is a solution to (\ref{eigrr}),(\ref{oplr}),(\ref{inc0}),(\ref{bou1}) for\\
$\tau\rme_\rr=(\Omega_1,\Omega_2,\Omega_3)$. Then (after an appropriate
permutation) the eight exponents $\gamma_s$ in (\ref{seriw}) satisfy
the relations
\begin{equation}\label{gamar3}
\renewcommand{\arraystretch}{1.3}
\begin{array}{c}
\gamma_1=k,\ \gamma_2=-k,\hbox{ and}\\
\hbox{either }\gamma_{2s+1}=-\bar\gamma_{2s+2}\hbox{ or }
\gamma_{2s+1}\hbox{ and }\gamma_{2s+2}\hbox{ are imaginary},\ s=1,2,3.
\end{array}
\end{equation}
For $s=3,...,8$ the exponents are roots of the equation
\begin{equation}\label{detr3}
(\gamma^2-k^2)^3+(\Omega_3\gamma-\ri\Omega_1k)^2+Rk^2=0
\end{equation}
and
\begin{equation}\label{dett3}
\det\cM=0,
\end{equation}
where the $8\times8$ matrix $\cM=(M_{ij})$ has the entries
\begin{equation}\label{matbr3}
\renewcommand{\arraystretch}{1.7}
\begin{array}{c}
M_{1,l}=-(\gamma_l^2-k^2){\rm e}^{\gamma_l/2},\
M_{2,l}=-(\gamma_l^2-k^2){\rm e}^{-\gamma_l/2},\\
M_{3,l}=-\gamma_l(\gamma_l^2-k^2){\rm e}^{\gamma_l/2},\
M_{4,l}=-\gamma_l(\gamma_l^2-k^2){\rm e}^{-\gamma_l/2},\\
M_{5,l}={\rm e}^{\gamma_l/2},\ M_{6,l}={\rm e}^{-\gamma_l/2},\\
M_{7,l}=-\displaystyle{(\gamma_l^2-k^2)^3+Rk^2\over
\Omega_3 k\gamma_l-\ri\Omega_1k^2}{\rm e}^{\gamma_l/2},\
M_{8,l}=-\displaystyle{(\gamma_l^2-k^2)^3+Rk^2\over
\Omega_3 k\gamma_l-\ri\Omega_1k^2}{\rm e}^{-\gamma_l/2}.
\end{array}
\end{equation}
\end{theorem}

\proof
We proceed as in the proofs of theorems \ref{th1} and \ref{th2}.
The subspace $F(k,\gamma)$ spanned by ${\bf f}_j(k,\gamma)$,
$j=1,2,3$, is $L$-invariant. For $\gamma\ne\pm k$, the matrix of $L$ acting on
$F(k,\gamma)$ is
\begin{equation}\label{matL3}
\left(
\renewcommand{\arraystretch}{1.5}
\begin{array}{ccc}
P(\gamma^2-k^2)&-\displaystyle\frac{P(\Omega_3 k\gamma-\ri\Omega_1k^2)}
{\gamma^2-k^2}&
-\displaystyle\frac{PRk^2}{\gamma^2-k^2}\\
\displaystyle{P(\Omega_3 \gamma-\ri\Omega_1k)\over k}&P(\gamma^2-k^2)&0\\
1&0&(\gamma^2-k^2)
\end{array}
\right).
\end{equation}
The operator possesses a non-trivial kernel whenever the determinant
of (\ref{matL3}) vanishes, i.e., when the equality (\ref{detr3})
holds true. For $\gamma=\pm k$,
\begin{equation}\label{detr33}
L[R{\bf f}_2(k,\gamma)+(\mp\Omega_3+\ri\Omega_1){\bf f}_3(k,\gamma)]=0.
\end{equation}

The change of variables $\gamma\to\ri\gamma$ turns (\ref{detr3}) into a
sixth-order algebraic equation with real coefficients. The modified equation
has either real roots, or pairs of complex-conjugate roots,
$\gamma$ and $\bar\gamma$.
Therefore, we can satisfy (\ref{gamar3}) by suitably numbering
the exponents $\gamma$.

By (\ref{matL3}) and (\ref{detr33}), if the sum (\ref{seriw}) is a solution to $L{\bf w}=0$, then
the coefficients $A_{jl}$ satisfy the following relations:
\begin{equation}\label{coeA3}
A_{1l}=-(\gamma^2-k^2)A_{3l},\quad
A_{2l}=-{(\gamma^2-k^2)^3+Rk^2\over \Omega_3 k\gamma-\ri\Omega_1k^2}A_{3l},\
l=1,\ldots,8.
\end{equation}

For a field (\ref{seriw}),(\ref{bas3}), boundary conditions (\ref{bou1}) reduce
to the relations
\begin{equation}\label{coeA33}
\sum_{l=1}^8 A_{jl}{\rm e}^{\pm\gamma_l/2}=0,\ j=1,2,3,\quad
\sum_{l=1}^8 A_{1l}\gamma_l{\rm e}^{\pm\gamma_l/2}=0.
\end{equation}
Due to (\ref{coeA3}), the equations (\ref{coeA33}) are a system of eight
linear equations in eight variables $A_{31},...,A_{38}$. The system involves
the matrix $\cM$ (\ref{matbr3}); it has a nontrivial solution whenever
(\ref{dett3}) is satisfied.
\qed

\begin{remark}\label{rm11}
For a convective layer rotating about a vertical axis (the case $\alpha=0$)
the $L$-invariant three-dimensional complex subspace spanned by
${\bf f}_1$, ${\bf f}_2$ and ${\bf f}_3$ is a union of two three-dimensional invariant
real subspaces. One of them is spanned by the real components of the vectors
${\bf f}_j$, the other one is spanned by the the imaginary components of the vectors.
The action of the operator $L$ on the first of these subspaces is considered in
theorem \ref{th1}. For a horizontal axis of rotation (the case $\alpha=\pi/2$)
the three-dimensional complex subspace also splits into a direct sum of two
$L$-invariant three-dimensional real subspaces. One of them is spanned by the real parts of  ${\bf f}_1$
and ${\bf f}_3$,  and the imaginary part of ${\bf f}_2$, the other subspace
is spanned by the imaginary parts of  ${\bf f}_1$
and ${\bf f}_3$,  and the real part of ${\bf f}_2$.
The operator $L$ acting on the first subspace is studied in theorem \ref{th2}.
\end{remark}

\begin{table}[ht]
\begin{tabular}{|c|c|}
\hline $\tau$ and $\alpha$ & $k$ \\ \hline
$\tau=100$, $\alpha=\pi/4$ & 4.3,5,6\\
$\tau=100$, $\alpha=\pi/8$ & 5,6,7\\
$\tau=500$, $\alpha=\pi/4$ & 7,8,9\\
$\tau=1000$, $\alpha=\pi/4$ & 9.6,12,14
\\\hline
\end{tabular}
\caption{Values of $\tau$, $\alpha$ and $k$ employed in
computation of $R_c(\tau,\alpha,k,\beta)$ with $0\le \beta\le \pi/2$.
}\label{tab1}
\end{table}

The theorem enables us to compute the critical Rayleigh number for
the monotonic onset of convection. For a fixed $\tau$ and $\alpha$, we denote
by $R_c(\tau,\alpha,k,\beta)$  the critical Rayleigh number, which is a solution
to the equations (\ref{detr3}), (\ref{dett3}), (\ref{matbr3}) for
$\bom=\tau(\sin\beta,\cos\beta\sin\alpha,\cos\alpha)$. The minimum
of $R_c(\tau,\alpha,k,\beta)$ over $k$ and $\beta$ is the critical Rayleigh number
for the onset of convection. It has been computed for several values of
$\tau$ and $\alpha$, see Table~\ref{tab1}. It turns out that for all the considered values
of $\tau$, $\alpha$ and $k$, the minimum of $R$ over $\beta$ is achieved for
$\beta=0$. Several instances of these computations are shown in Fig.~1.
Thus, like for the horizontal rotation axis, the parallel mode is the most
unstable. It remains desirable to find an analytical proof of this result.
Consequently, the critical Rayleigh number is
known from the studies of convection with the vertical rotation axis.

\centerline{
\psfig{file=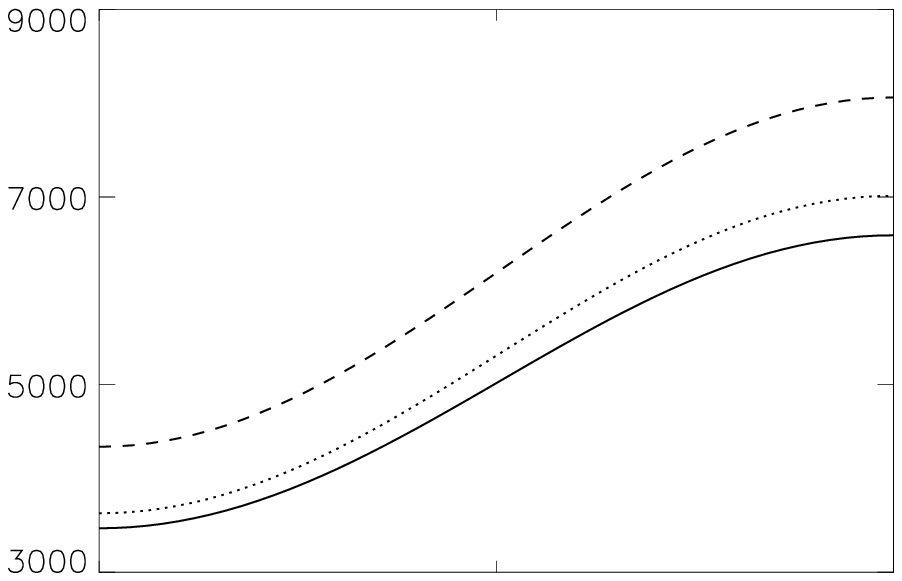,width=9cm}~\hspace{-2mm}
\psfig{file=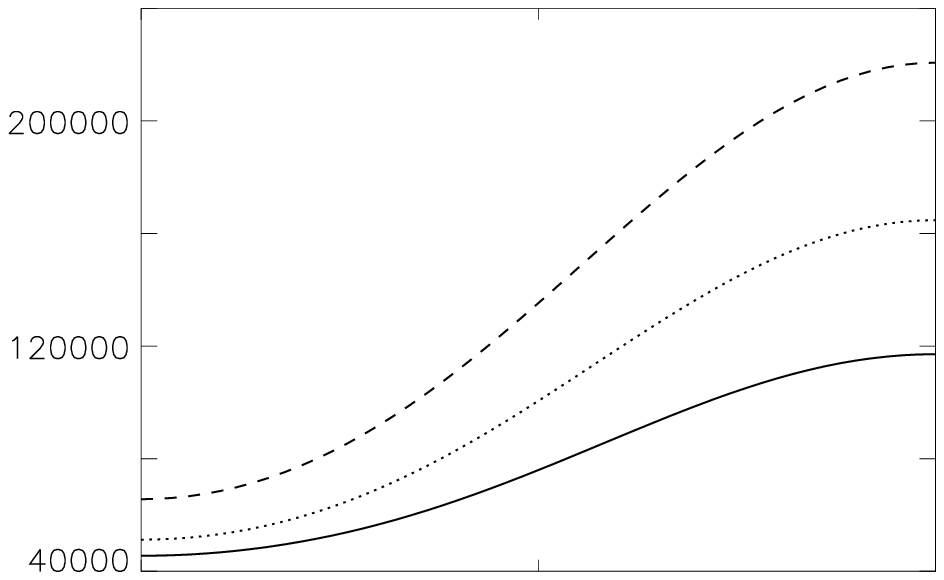,width=9cm}
}

\vspace*{-.9cm}
\hspace*{1mm}
{ 0\hspace*{60mm}$\pi/2$\hspace*{22mm}0\hspace*{60mm}$\pi/2$}

\vspace*{-.1cm}
\hspace*{3.5cm}
{\large $\beta$\hspace*{87mm}$\beta$}

\vspace*{-40mm}
\hspace*{-12mm}
{\large $R$\hspace*{86mm}$R$}

\vspace{37mm}
\centerline{
(a)\hspace{86mm}(b)
}

\medskip\noindent
Figure 1.
Critical Rayleigh number $R_c(\tau,\alpha,k,\beta)$ as a function of $\beta$
for $\alpha=\pi/4$ and\\
(a) $\tau=100$, $k=4.3$ (solid line), $k=5$
(dotted line) or $k=6$ (dashed line); (b) $\tau=1000$, $k=9.6$ (solid line),
$k=12$ (dotted line) or $k=14$ (dashed line).

\section{Nonlinear convection}\label{sec6}

For small $R$ the fluid is not moving and the heat is transported by the
thermal diffusion only. As the Rayleigh number exceeds the critical value, the
fluid motion sets in.
As shown in sections \ref{sec4} and \ref{sec5} for
a layer rotating with the angular velocity $(0,\Omega_2,\Omega_3)$ the most
unstable mode corresponds to rolls aligned with the $y$-axis.
By theorem \ref{th0} the solution to the nonlinear problem
(\ref{nst})--(\ref{inc0}) emerging from this mode is independent of
$\Omega_2$ (up to a modification of the pressure).
However, the theorem tells us nothing about its stability.
In this section we study
numerically how the stability of this primary solution, convective
rolls, varies as $\Omega_2$ is increased.
Namely, we construct bifurcation diagrams for a range of $R$, several
values of $\tau$ and $P=1$. We consider the case of horizontal
rotation axis, $\rme_\rr=(0,1,0)$, and an inclined axis with
$\Omega_3=250$ and four values of $\Omega_2$.

We consider solutions, periodic in the horizontal variables, the size
of a square periodicity cell being related to the critical wave number.
The conditions (\ref{bou1}) are assumed on the horizontal boundaries.
We expand the fields in trigonometric series in $x$ and $y$ and Chebyshev
polynomials in $z$, employ pseudospectral methods (see, e.g., \cite{canu}).
For integration in time we apply a fourth-order Runge--Kutta scheme. More
details about the method will be given in a forthcoming paper \cite{op23}.

Let us compare the critical values for the onset of convection obtained by
different methods.
According to \cite{chan}, in a non-rotating layer the critical value is
$R^{\cri}=1707.762$ and the respective critical wave number is $k^{\cri}=3.117$.
Solving numerically equations (\ref{detr00}) and (\ref{matbr0}) for
$\Omega_3=0$ and minimising over $k$ (see section \ref{sec3}), we have obtained
$R^{\cri}=1707.8$ and $k^{\cri}=3.1$. Numerical integration of
equations (\ref{nst})--(\ref{inc0}) for $\Omega_3=0$ and $k=3.1$ (see Fig.~2)
yields the onset of the convective motion between $R=1708$ and $R=1709$
for all considered $\Omega_2$.

In a layer rotating about a vertical axis, for $\Omega_3=250$
the critical values found numerically by solving equations
(\ref{detr00}) and (\ref{matbr0}) and minimising over $k$
(see section \ref{sec3}) are $R^{\cri}=12648$ and $k^{\cri}=6.6$. The same (to this
precision) critical values have been found for this $\Omega_3$ and
several values of $\Omega_2$ from equations
(\ref{detr3}), (\ref{dett3}) and (\ref{matbr3}) (section \ref{sec5}).
The critical values coincide because for $\Omega_2=0$
equation (\ref{detr3}) turns into (\ref{detr00}), and equations (\ref{dett3})
and (\ref{matbr3}) into (\ref{matbr0}).
Numerical integration of equations (\ref{nst})--(\ref{inc0}),
see Fig.~4, has shown that $(\bv=0,\theta=0)$ is stable for $R=12656$
and unstable for $R=12657$. Thus, the difference in the critical values
of the Rayleigh number is below 0.1\%.

The computed bifurcation diagrams for $P=1$, $0<R\le50000$, and
$\tau=0,250,500$ and 1000 are shown in Fig.~2. As expected, when the Rayleigh
number exceeds the critical value for the onset of the non-rotating convection,
the fluid motion in the form of rolls sets in. For $\tau=0$, at $R\approx21000$
the rolls become unstable and a travelling wave (see.Fig.~3,a,b) emerges
in a supercritical bifurcation. (A travelling wave is a flow that is steady
in a reference frame moving with the pattern, and time-periodic in a non-moving
one.) On increasing $R$, the behaviour becomes quasiperiodic and afterwards
chaotic (we do not study the respective bifurcations in detail).

For $\tau=250$, the interval of the stability of rolls becomes smaller,
the supercritical bifurcation to wavy rolls (WR, i.e., a steady flow
in the form of deformed rolls; see Fig.~3c) occurs at $R\approx13000$, and
WR become unstable in a Hopf bifurcation at $R\approx20000$. When
$\tau$ is increased from 250 to 500 and further to 1000, we observe that
the interval of the stability of rolls is growing: for $\tau=500$,
the bifurcation to WR takes place at $R\approx33000$, and for $\tau=1000$
the rolls are stable up to the maximum $R$ employed in computations.
It is desirable to find the asymptotics of the critical Rayleigh number
for the instability of rolls in $\tau$ in the limit of large $\tau$.

\pagebreak

\centerline{
\psfig{file=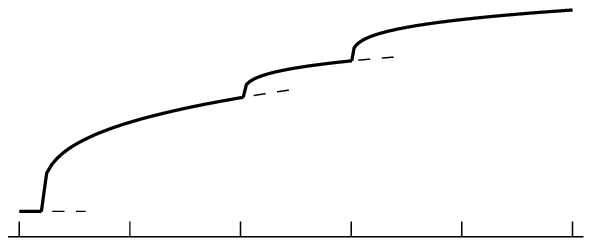,width=11cm}~\hspace{-22mm}
\psfig{file=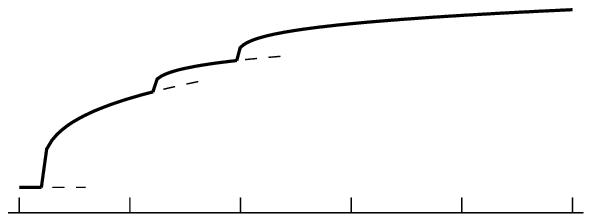,width=11cm}
}

\vspace*{-1.9cm}
\hspace*{1mm}
{ 0\hspace*{60mm}50000\hspace*{18mm}0\hspace*{60mm}50000}

\hspace*{33mm}
{\large $R$\hspace*{87mm}$R$}

\vspace*{-2.9cm}
\hspace*{.6cm}
{rolls\hspace*{83mm}rolls}

\vspace*{-1.2cm}
\hspace*{28mm}TW

\vspace*{-.3cm}
\hspace*{108mm}WR

\vspace*{-1.4cm}
\hspace*{5.4cm}QP,C

\vspace*{-.3cm}
\hspace*{14.6cm}P,QP,C

\vspace{39mm}
\centerline{
(a)\hspace{80mm}(b)
}

\centerline{
\psfig{file=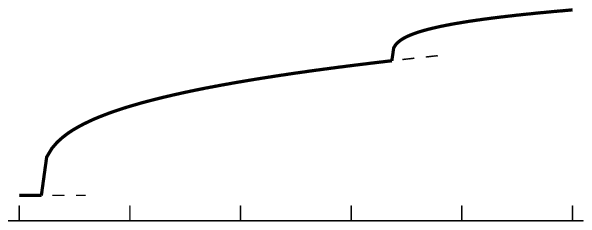,width=11cm}~\hspace{-22mm}
\psfig{file=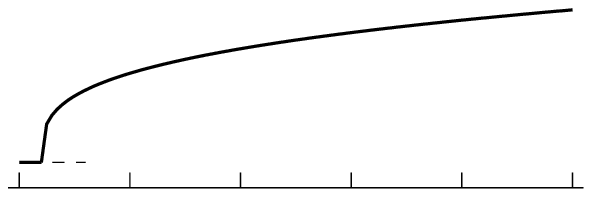,width=11cm}
}

\vspace*{-1.9cm}
\hspace*{1mm}
{ 0\hspace*{60mm}50000\hspace*{18mm}0\hspace*{60mm}50000}

\hspace*{33mm}
{\large $R$\hspace*{87mm}$R$}

\vspace*{-2.9cm}
\hspace*{.6cm}
{rolls\hspace*{83mm}rolls}

\vspace*{-1.6cm}
\hspace*{44mm}WR

\vspace{39mm}
\centerline{
(c)\hspace{80mm}(d)
}

\medskip\noindent

Figure 2.
Schematic bifurcation diagrams for a convective layer rotating about a horizontal axis
for $P=1$, $k=3.1$, $\rme_\rr=(0,1,0)$ and (a) $\tau=0$; (b) $\tau=250$; (c) $\tau=500$
and (d) $\tau=1000$.

\pagebreak
\centerline{
\psfig{file=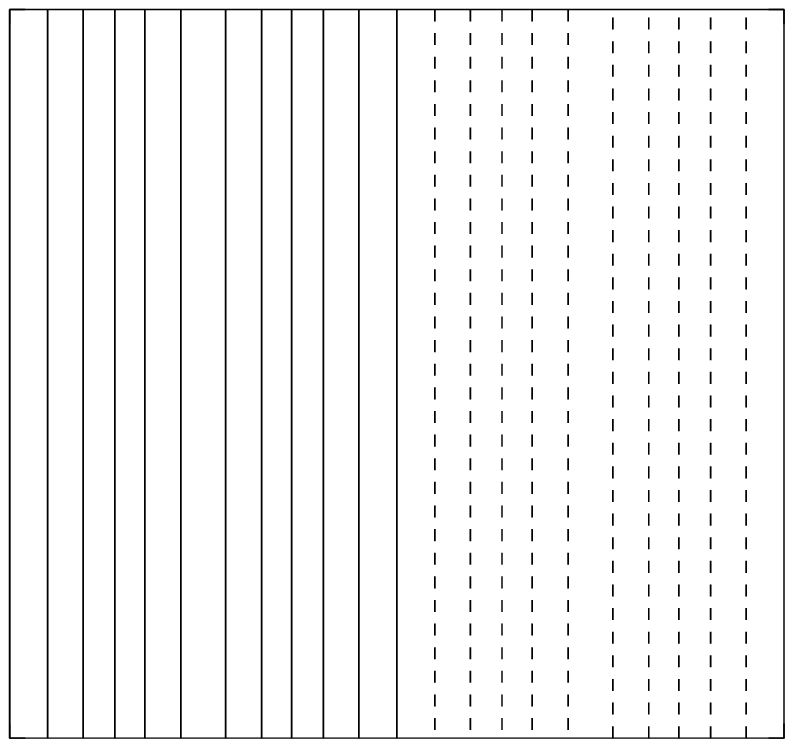,width=5cm}~\hspace{4mm}
\psfig{file=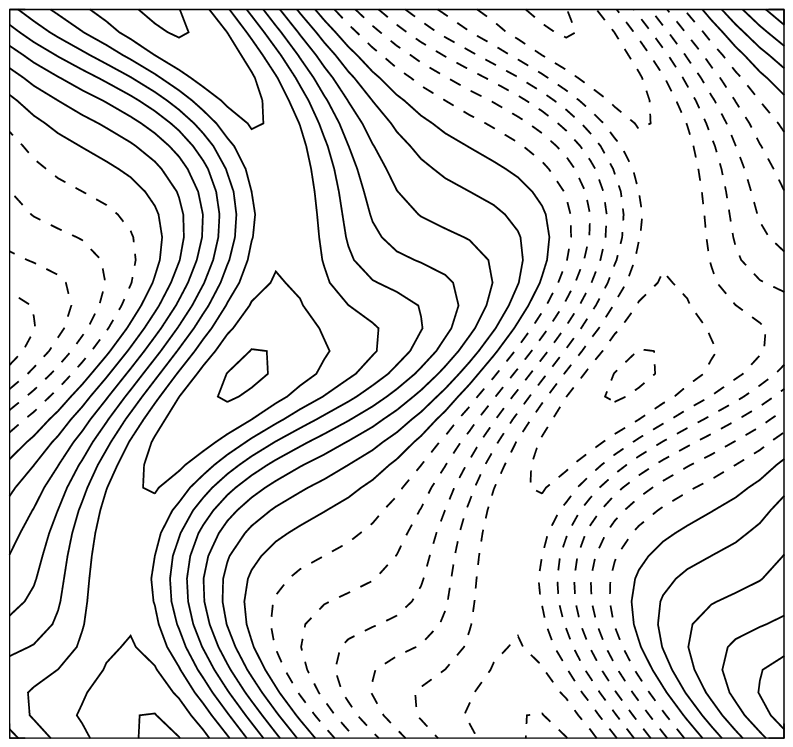,width=5cm}~\hspace{4mm}
\psfig{file=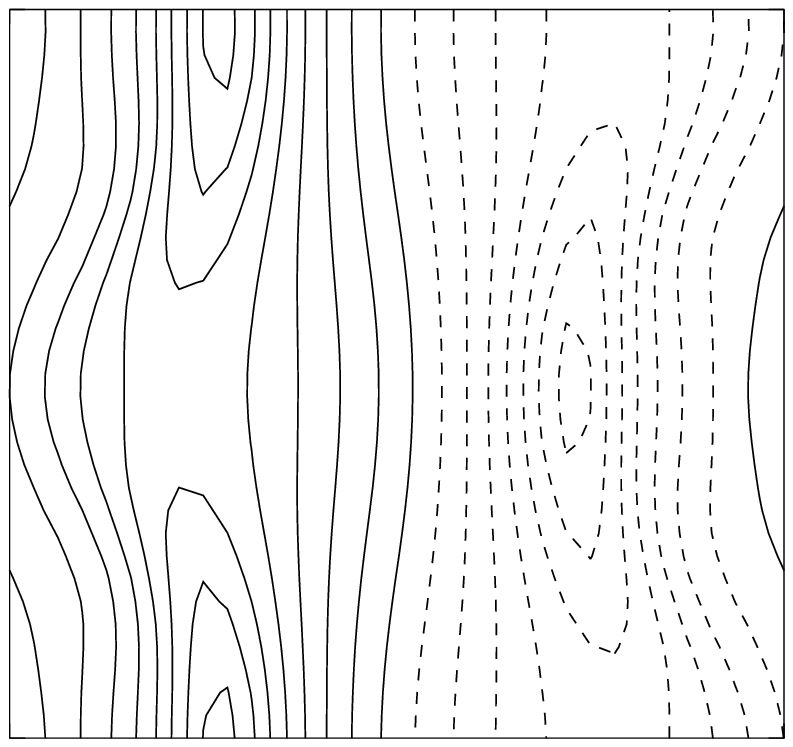,width=5cm}
}

\vspace*{-.4cm}
{ 0\hspace*{38mm}L\hspace*{14.5mm}0\hspace*{38mm}L\hspace*{14.5mm}0\hspace*{38mm}L}

\vspace*{-1.1cm}
\hspace*{-4mm}
{$0$\hspace*{55mm}$0$\hspace*{55mm}$0$}

\vspace*{-3.9cm}
\hspace*{-.4cm}
{L\hspace*{53mm}L\hspace*{53mm}L}

\vspace*{3.7cm}
\hspace*{1.9cm}
{\large $x$\hspace*{54.5mm}$x$\hspace*{54.5mm}$x$}

\vspace*{-3.1cm}
\hspace*{-.7cm}
{\large $y$\hspace*{54.5mm}$y$\hspace*{54.5mm}$y$}

\vspace{30mm}
\centerline{
(a)\hspace{53mm}(b)\hspace{53mm}(c)
}

\medskip\noindent
Figure 3.
Isolines (step 10) of the vertical component of the velocity in the horizontal
midplane for $P=1$, $\rme_\rr=(0,1,0)$ and (a) $\tau=0$, $R=20000$ (rolls),
(b) $\tau=0$, $R=30000$ (TW) and $\tau=500$, $R=50000$ (WR).

\vspace*{1.cm}

The results of numerical simulations of convective flows in a layer rotating
about an inclined axis are summarized as bifurcation diagrams shown in Fig.~4.
The computations have been performed for several values of $\Omega_2$, other
parameters being fixed.
remains a solution for any $\Omega_2$.
We observe a monotonic growth
of the interval in $R$ of the stability of rolls as $\Omega_2$ increases,
similarly to the case of the layer rotating about a horizontal axis.

\section{Conclusion}

We have studied the monotonic onset of convection and the stability of emerging convective
rolls in a horizontal layer rotating about an inclined axis with the angular
velocity $(\Omega_1,\Omega_2,\Omega_3)$. At the onset of convection,
the critical Rayleigh number, the critical wave number and the emerging
convective rolls (two-dimensional flows) coincide with those in the layer
rotating about the vertical axis with the angular velocity $\Omega_3{\bf e}_3$.
Numerical simulations performed for $\Omega_1=0$, fixed $\Omega_3=0$ and 250,
several values of $\Omega_2$ and a range of $R$ ($0\le R\le5\cdot10^4$ for
$\Omega_3=0$ and $0\le R\le10^5$ for $\Omega_3=250$) show that the convective
rolls become more stable on increasing $\Omega_2$: the interval of $R$,
where rolls are stable, in general, increases with $\Omega_2$. It remains
desirable to establish the asymptotics in $\Omega_2$ of the critical Rayleigh
number for the instability of rolls in the limit $\Omega_2\to\infty$.

\pagebreak

\centerline{
\psfig{file=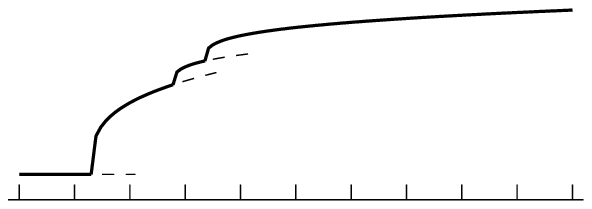,width=11cm}~\hspace{-22mm}
\psfig{file=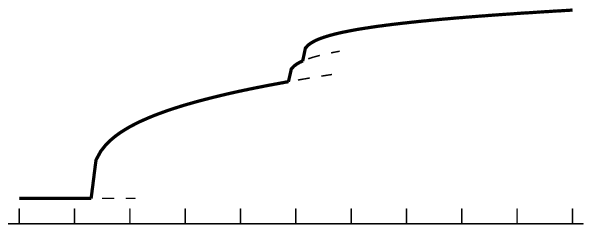,width=11cm}
}

\vspace*{-1.9cm}
\hspace*{1mm}
{ 0\hspace*{60mm}$10^5$\hspace*{23mm}0\hspace*{60mm}$10^5$}

\hspace*{33mm}
{\large $R$\hspace*{87mm}$R$}

\vspace*{-2.7cm}
\hspace*{.6cm}
{rolls\hspace*{83mm}rolls}

\vspace*{-9mm}
\hspace*{15mm}WR

\vspace*{-7mm}
\hspace*{118mm}WR

\vspace*{-1.1cm}
\hspace*{5.4cm}P,QP,C

\vspace*{-8mm}
\hspace*{14.6cm}P,QP,C

\vspace{39mm}
\centerline{
(a)\hspace{80mm}(b)
}

\centerline{
\psfig{file=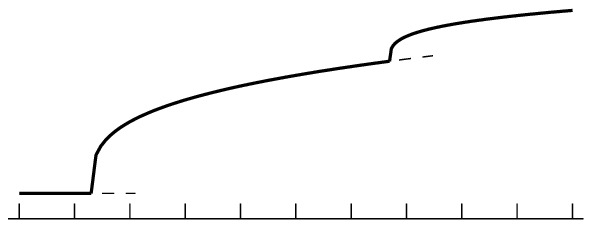,width=11cm}~\hspace{-22mm}
\psfig{file=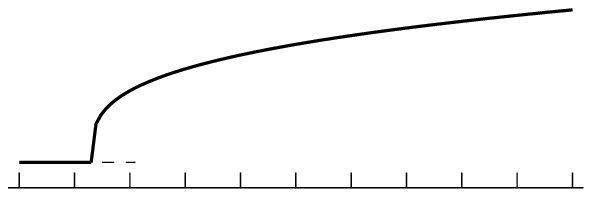,width=11cm}
}

\vspace*{-1.9cm}
\hspace*{1mm}
{ 0\hspace*{60mm}$10^5$\hspace*{23mm}0\hspace*{60mm}$10^5$}

\hspace*{33mm}
{\large $R$\hspace*{87mm}$R$}

\vspace*{-2.7cm}
\hspace*{.6cm}
{rolls\hspace*{83mm}rolls}

\vspace*{-1.6cm}
\hspace*{42mm}WR

\vspace{37mm}
\centerline{
(c)\hspace{80mm}(d)
}

\medskip\noindent

Figure 4.
Schematic bifurcation diagrams in rotating convection for $P=1$, $k=6.6$, $\Omega_1=0$, $\Omega_3=250$ and\\
(a) $\Omega_2=0$; (b) $\Omega_2=250$; (c) $\Omega_2=500$; (d) $\Omega_2=1000$.

\vspace*{1.cm}

At present, magnetic fields of stars and planets are supposed to be generated
by convection. Consequently, convective flows are often used as a flow velocity
field in the kinematic dynamo problem in geophysical and astrophysical
applications. We have proven that the emerging convective rolls are independent
of the horizontal part of the angular velocity, $\Omega_1$ and $\Omega_2$.
Thus, results on magnetic field generation by rolls in a fluid layer rotating
about a vertical axis (see, e.g., \cite{math,che}) also pertain for rolls
in a layer rotating about an inclined axis. In particular, while rolls
in a layer rotating about a horizontal axis do not generate magnetic fields,
for the inclined axis this is possible.

In this paper, the onset of convection has been studied following the ideas of
Pellew and Southwel \cite{ps40}, which were also applied in \cite{kno1,zhan3,op10}
 for more involved convective problems.
The method is likely to be useful for investigating a large number of problems,
including convection in an inclined layer, convection with inclined rotation
and imposed magnetic field, convection of compressible flows, convection in a
rotating layer with shear (see, e.g., Ponty\al\ \cite{sow}).

\bigskip
{\large\bf Acknowledgments}
The project was financed by the grant
{\fontfamily{cmr}\selectfont\textnumero}\,22-17-00114 of the Russian
Science Foundation, https://rscf.ru/project/22-17-00114/.

\bibliographystyle{apsrev}

\begin{thebibliography}{99}

\bibitem{gla}
G.A. Glatzmaier and P.H. Roberts,
A Three-Dimensional Self-Consistent Computer Simulation of a Geomagnetic Field
Reversal. {\it Nature}, {\bf 377}, 203-209; (1995).

\bibitem{rob}
P.H. Roberts and E.M. King On the genesis of the Earth's magnetism.
{\it Rep. Prog. Phys.} {\bf 76}, 096801 (2013).

\bibitem{chr}
U.R. Christensen and J.  Wicht, Numerical dynamo simulations.
{\it Treatise on Geophysics} {\bf 8}, 245-277 (2015).

\bibitem{chan}
S. Chandrasekhar, {\it Hydrodynamic and Hydromagnetic Stability.} (Oxford,
Clarendon Press, 1961).

\bibitem{bus}
F.H. Busse,
Non-linear properties of thermal convection.
{\it Rep. Prog. Phys.}, {\bf 41}, 1929-1967, (1968).

\bibitem{get}
A.V. Getling,
{\it Rayleigh-B\'enard Convection: Structures and Dynamics}.
World Scientific, 245pp., (1998).

\bibitem{kno1}
T. Clune and E. Knobloch,
Pattern selection in rotating convection with experimental boundary conditions.
{\it Phys. Rev. E.} {\bf 4}, 2536, (1993).

\bibitem{kl69}
G. K\"uppers and D. Lortz, J.~Fluid Mech.
Transition from laminar convection to thermal turbulence in a rotating fluid layer.
 {\it J.~Fluid Mech.} {\bf 35}, 609-620, (1969).

\bibitem{cb79}
R.M. Clever and F.H. Busse,
Nonlinear properties of convection rolls in a horizontal layer rotating about a vertical axis.
{\it J.~Fluid Mech.} {\bf 94}, 609, (1979).

\bibitem{bpa}
E. Bodenschatz, W. Pesch and G. Ahlers,
Recent developments in Rayleigh-B\'enard convection.
{\it Ann. Rev. Fluid Mech.} {\bf 32}, 709, (2000).

\bibitem{kba02}
K.M.S. Bajaj, G. Ahlers and W. Pesch,
Rayleigh-B\'e
nard convection with rotation at small Prandtl numbers.
{\it Phys. Rev. E.} {\bf 65}, 056309, (2002).

\bibitem{htg80}
D.H. Hathaway, J. Toomre and P.A. Gilman,
Convective instability when the temperature gradient and rotation vector are
oblique to gravity. II. Real fluids with effects of diffusion.
{\it Geophys. Astrophys. Fluid Dyn.} {\bf 15}, 7-37, (1980).

\bibitem{hs83}
D.H. Hathaway and R.C. Somerville, Three-dimensional simulations of convection
in layers with tilted rotation vectors.
{\it J.~Fluid Mech.} {\bf 126}, 75-89, (1983).

\bibitem{jk98}
K. Julien and E. Knobloch,
Strongly nonlinear convection cells in a rapidly
rotating fluid layer: the tilted $f$-plane.
{\it J.~Fluid Mech.} {\bf 360}, 141-178, (1998).

\bibitem{el72}
I.A. Eltayeb,
Hydromagnetic convection in a rapidly rotating fluid layer.
{\it Proc. Roy. Soc. (London) A} {\bf 326}, 229, (1972).

\bibitem{el75}
I.A. Eltayeb,
Overstable hydromagnetic convection in a rotating fluid layer.
{\it J.~Fluid Mech.} {\bf 71}, 161, (1975).

\bibitem{zhan3}
K. Zhang, M. Weeks and P. Roberts,
Effect of electrically conducting walls on rotating magnetoconvection.
{\it Phys. Fluids} {\bf 16}, 2023, (2004).

\bibitem{op10}
O. Podvigina,
Stability of rolls in rotating magnetoconvection in a layer with no-slip electrically insulating horizontal boundaries.
{\it Phys. Rev. E}, {\bf 81}, 056322, (2010).

\bibitem{sow}
Y. Ponty, A.D. Gilbert and A.M. Soward,
Kinematic dynamo action in large magnetic Reynolds number flows driven by shear and convection.
{\it JFM}, {\bf 435} , 261 - 287, (2001).

\bibitem{ps40}
A. Pellew and R. V. Southwell,
On the maintained convective motion in a fluid heated from below.
{\it  Proc. Roy. Soc. (London) A} {\bf 176}, 312, (1940).

\bibitem{zhan1}
M. Weeks and K. Zhang,
Thermal generation of Alfv\'en waves in oscillatory magnetoconvection: diffusively modified modes.
{\it Geophys. Astrophys. Fluid Dyn.} {\bf 96}, 405-424, (2002).

\bibitem{canu}
C. Canuto, M. Hussaini, A. Quarteroni, T.A. Zang
{\it Spectral Methods in Fluid Dynamics.} Springer, (1988).

\bibitem{op23}
O. Podvigina,
An efficient Galerkin method for problems with physically
realistic boundary conditions. Submitted to {\it JCP}.

\bibitem{math}
P.C. Matthews,
Dynamo action in simple convective flows,
{\it Proceedings of the Royal Society of London. Series A}, {\bf 455}, 1829-1840, (1999).

\bibitem{che}
R. Chertovskih, S.M.A. Gama, O. Podvigina and V. Zheligovsky,
Dependence of magnetic field generation by thermal convection on the rotation
rate: A case study, {\it Physica D} {\bf 239}, 1188-1209, (2010).
\end{thebibliography}

\end{document}